\documentclass[twocolumn]{aastex62}

\usepackage[utf8]{inputenc}

\submitjournal{ApJ}

\shorttitle{IMBH in the A85 Dwarf Galaxy}
\shortauthors{Mićić et al.}
\graphicspath{{./}{figures/}}
\bibpunct[, ]{(}{)}{;}{a}{,}{,}
\begin{document}

\title{An Ultraluminous Supersoft Source in a Dwarf Galaxy of Abell 85: An Intermediate-Mass Black Hole Candidate}

\correspondingauthor{Marko Mićić}
\email{mmicic@crimson.ua.edu}

\author{Marko Mićić}
\affiliation{Department of Physics \& Astronomy, University of Alabama, Tuscaloosa, AL 35401, USA}

\author{Jimmy A. Irwin}
\affiliation{Department of Physics \& Astronomy, University of Alabama, Tuscaloosa, AL 35401, USA}
\author{Dacheng Lin}
\affiliation{Department of Physics, Northeastern University, Boston, MA 02115-5000, USA}


\begin{abstract}

We study a large sample of dwarf galaxies using archival \textit{Chandra} X-ray observations, with the aim of detecting accreting intermediate-mass black holes (IMBHs). IMBHs are expected to inhabit dwarf galaxies and to produce specific signatures in terms of luminosity and X-ray spectra. We report the discovery of an X-ray source associated with an Abell 85 dwarf galaxy that fits the IMBH description. The stellar mass of the host galaxy is estimated to be 2 $\times$ 10$^8$ \(\textup{M}_\odot\), which makes it one of the least massive galaxies to potentially host an accreting black hole. The source is detected in the soft band, under 1 keV, while undetected at higher energies. The X-ray luminosity is $\approx$ 10$^{41}$ erg s$^{-1}$, making it almost three orders of magnitude more luminous than the most luminous stellar-mass supersoft emitters. From the galaxy stellar mass vs.\ black hole mass relation, we estimate the mass to be within the intermediate regime. Another method that resulted in an intermediate mass relies on the fact that supersoft emission is expected to be associated with high accretion rates, approaching the Eddington limit. We suggest that the observed offset of the X-ray source from the galactic center ($\approx$ 1.8 kpc) is due to galaxy interactions, and we present evidence from the literature that supports the relation between black hole activity and galaxy interactions. 

\end{abstract}

\keywords{galaxy clusters -- dwarf galaxies -- black holes}


\section{Introduction} \label{sec:intro}
Observational evidence supports the existence of two types of black holes: (1) stellar-mass black holes, which form as end-products of the life of the most massive stars, with masses below 10$^2$ \(\textup{M}_\odot\) and (2) supermassive black holes (SMBHs), with masses exceeding 10$^5$ \(\textup{M}_\odot\), which are located in the centers of galaxies. An open question in astronomy is how black holes evolve and reach supermassive sizes. One of the postulated mechanisms is that massive black holes with masses 10$^4$--10$^6$ \(\textup{M}_\odot\) are formed in the early Universe through the process of direct collapse of massive gas clouds \citep{1994ApJ...432...52L, Bromm_2003}. Another suggests that SMBHs emerge from low-mass seeds, black holes of M$_\mathrm{BH}\approx$10$^2$ \(\textup{M}_\odot\), the end-products of the first, very massive Population III stars \citep{Madau_2001}. However, the latter scenario would leave behind a significant population of black holes with masses in between the stellar and supermassive regime, 10$^2$--10$^5$ \(\textup{M}_\odot\), the so-called intermediate-mass black holes (IMBHs). Although substantial effort has been spent searching for IMBHs, the evidence for their existence remains scarce. \newline \indent
From many lines of reasoning, IMBHs are expected to inhabit dwarf galaxies. First, the relation between the stellar mass of the galaxy and black hole mass ($M_*-M_\mathrm{BH}$) clearly indicates that the least massive black holes are expected to be associated with the least massive galaxies \citep{Reines_2015}. Also, dwarf galaxies undergo a quieter merger history, and therefore it is expected for their central black hole to preserve the information about the initial seed. \newline \indent
However, the existence of IMBHs in dwarf galaxies and other low-stellar mass systems, such as globular clusters, is still controversial. For example, dynamical measurements of stars in central regions of many globular clusters resulted in a few IMBH candidates \citep{2010ApJ...719L..60N}, but competing physical mechanisms and limitations in available data make these results debatable \citep{10.1093/mnras/stx316}. Also, gas and stellar dynamics measurements are unfeasible at larger distances in the case of low-mass black holes in low-mass galaxies since the gravitational sphere of influence cannot be resolved beyond a few Mpc. For example, the sphere of influence of a 10$^5$ \(\textup{M}_\odot\) at a distance of 10 Mpc is $\approx$ 0\farcs01 \citep{2018ApJ...861..142C}. A more powerful method that can help us identify black holes focuses on detecting the signatures of black hole accretion. Indeed, recently an increasing number of accreting black holes in low-mass galaxies have been detected using X-ray observatories \citep{Lemons_2015, Pardo_2016, birc10.1093/mnras/staa040}. \newline \indent
Accreting IMBHs are expected to produce specific signatures in terms of X-ray luminosity and spectrum. Emission from an AGN in the high state is thought to be dominated by thermal emission from a geometrically thick and optically thin accretion disk \citep{1973A&A....24..337S}. High states are associated with high mass accretion rates, and an accretor in this state will emit near its Eddington luminosity ($>$10\% L$_{Edd}$), where the Eddington bolometric luminosity is given as L$_{Edd}$=1.4$\times$ 10$^{38}$ M/\(\textup{M}_\odot\) erg s$^{-1}$. The high states allow us to estimate the mass of the accretor since the temperature of the accretion disk scales with the accretor's mass as T$\propto$ M$_{\textup{BH}}^{-1/4}$. From this relation, it follows that with increased mass, the temperature of the accretion disk decreases. Therefore, a typical stellar-mass black hole would be characterized by a color temperature of 1-2 keV, while in the case of SMBHs, this temperature would be in the UV range. On the other hand, an IMBH would have a color temperature below 0.5 keV, but still in the X-ray regime \citep{Liu2008}. As an example, a typical 10$^4$ \(\textup{M}_\odot\) IMBH, in its high state, should have a luminosity of L$_X$=10$^{40}$--10$^{41}$ erg s$^{-1}$, and emission confined below 1 keV. A SMBH could account for L$_X\approx$ 10$^{41}$ erg s$^{-1}$, but this could only occur in its low, non-thermal state, which should not produce a supersoft X-ray emission signature. Stellar-mass black holes cannot account for such high luminosities, low disk temperatures and supersoft X-ray emission. Some of the well known IMBH candidates with supersoft spectra include 3XMM J215022.4-055108 \citep{2020ApJ...892L..25L}, 3XMM J141711.1+522541 \citep{2016ApJ...821...25L} and ESO 243-49 HLX-1 \citep{2011ApJ...743....6S}.\newline \indent
Having this in mind, we have begun to carry out a large-scale study of archival \textit{Chandra} X-ray observations of dwarf galaxies in order to search for high luminosity, supersoft X-ray sources, and therefore provide compelling evidence for the existence of IMBH in such galaxies. In Section 2 we present optical and X-ray data relevant for this paper and data reduction methods. In Section 3 we present our results, and in Section 4 we discuss the nature and properties of our detection. 
\section{Data sample and analysis} \label{sec:data}
    This paper is a part of a broader research, in which we crossmatched galaxies from various catalogs with archival \textit{Chandra} observations to build a sample of $\approx$ 60,000 dwarf galaxies, with deep X-ray coverage. Specifically, for the case of Abell 85, after surveying the \textit{Chandra} database, we found four overlapping ACIS-I observations, with a total exposure time of 156.82 ksec. Observations were taken in 2013, within ten days (PI: Allen). For X-ray data reduction we used \textbf{CIAO 4.12} tools, and Calibration Database \textbf{CALDB 4.9.3}. All observations were reprocessed using the \textbf{chandra\_repro} tool. We used the \textbf{merge\_obs} tool to reproject and combine individual observations to create a merged event file and exposure-corrected images. From the merged event file, we constructed images in soft (0.3--1 keV) and hard (1--6 keV) bands, which allowed us to detect the presence of supersoft X-ray sources. Also, as we already mentioned, a 10$^4$ \(\textup{M}_\odot\) IMBH, in its high state, should have a luminosity of $\approx$ 10$^{40}$-10$^{41}$ erg s$^{-1}$, and our X-ray data is sufficiently deep to detect sources of such luminosities even in the regions of Abell 85 with strong X-ray emission coming from the hot intra-cluster gas. A summary of used observations is given in Table 1.
        \begin{table}
  \centering
  \begin{tabular}{|c|c|c|c|}
  \hline
    Obs.ID & Obs. Date & Exp. time &Off-axis angle\\
    & & (ksec)&\\
    \hline
    15173  & 2013-08-14&42.52&3\farcm124\\
    15174&2013-08-09&39.55&7\farcm010\\
    16263 &2013-08-10&38.15&7\farcm010\\
    16264& 2013-08-17&36.60&3\farcm125\\
    \hline
  \end{tabular}
  \caption{Summary of \textit{Chandra} ACIS-I observations used in this paper.}
  \label{tab:1}
\end{table}
\newline \indent
Prior to merging the observations we performed astrometric corrections on each observation. We started by running the CIAO tool wavdetect across the full Abell 85 field of view in order to detect X-ray sources. The parameters were set in such way to neglect small, faint sources and focus only on strong sources with robust centroids. Next, we matched \textit{Chandra} sources with an external catalog, SDSS DR7. The CIAO tool wcs$\_$match takes in previously selected X-ray and SDSS sources, iteratively crossmatches them, and determines transformation parameters in order to minimize positional differences. Next, the CIAO tool wcs$\_$update is used to apply the correction parameters and update the \textit{Chandra} event and aspect solution files. The procedure was repeated for every individual observation and only then updated and corrected event files are merged into one. The astrometric corrections procedure was also performed by hand, in order to check the correctness of the previous method. We used a GAIA source with a strong X-ray counterpart and calculated the offset between the object's position and X-ray source centroid. The offset was then applied to \textit{Chandra} event files, resulting in the same outcome as our initial approach.

\begin{figure}[t]
\epsscale{1}
\plotone{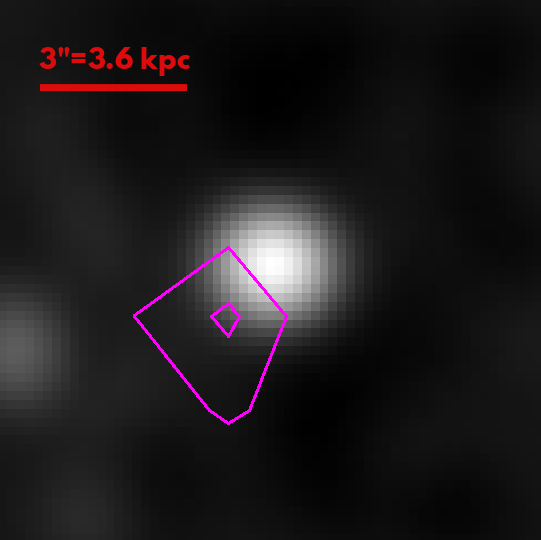}
\caption{\textit{Canadian-French-Hawaiian Telescope} G band image of WJ004 dwarf galaxy with overlaid magenta coloured X-ray contours corrected for astrometry.
\label{fig:fig3}}
\end{figure}

\begin{figure*}[t]
\epsscale{1}
\plotone{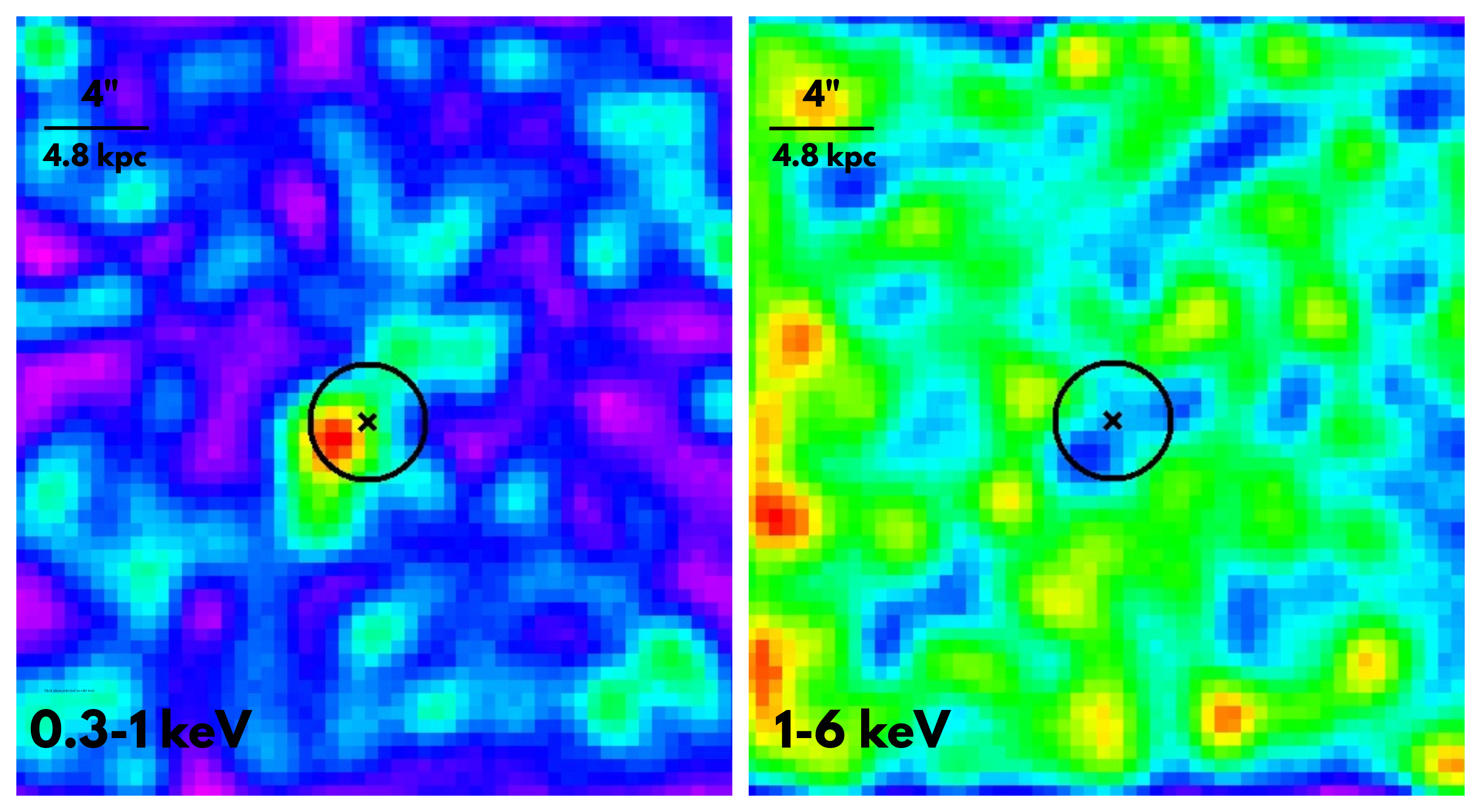}
\caption{Merged \textit{Chandra} X-ray images of the supersoft source in the soft band (left) and the hard band (right). Images are smoothed with a Gaussian with a kernel with a radius of three. The black $\times$ symbol represents the optical center of the host galaxy, WJ004, and the black circle represents the galaxy's Kron radius of 2\farcs2, as given in \cite{varela}. Diffuse intra-cluster medium X-ray emission is visible in the hard band image.
\label{fig:fig1}}
\end{figure*}
    In order to build a sample of the Abell 85 dwarf galaxies, we used a WIde-field Nearby Galaxy-cluster Survey (WINGS) catalog \citep{varela}. Using the data from the WINGS catalog, and assuming that galaxies are located at the distance of Abell 85 ($z$=0.0555) \citep{1998A&AS..129..281D}, we computed the V-band absolute magnitudes (M$_V$). To select the smallest and faintest, genuine dwarfs, we kept only galaxies fainter than M$_V$=--14. Next, as a final filter, we plotted selected dwarfs over the stacked X-ray field of view, which resulted in 220 dwarf galaxies fainter than  M$_V$=--14 with X-ray coverage. In the final step of data preparation, we cut out small (30\arcsec$\times$30\arcsec) X-ray images, centered on each of 220 dwarfs, in both soft and hard bands. 
\section{Results} \label{sec:results}
\subsection{The host galaxy}
Applying the above procedure to our sample of dwarf galaxies resulted in detecting one supersoft, luminous X-ray source associated with the galaxy WINGS J004144.10-091726 (WJ004 hereafter). WJ004 is a faint dwarf galaxy, with apparent magnitudes m$_V$=23.15 and m$_B$=24.3. Adopting a distance of 249 Mpc, its absolute magnitudes are M$_V$=--13.8 and M$_B$=--12.7. We calculated a V-band-mass-to-light ratio using the $B-V$ color index and following the equation log$ (M/L_V)$=$a_V$+($b_V$ $\times$ color), where $a_V$=--0.628 and $b_V$=1.305, as given in \cite{Bell_2003}. From this, we computed the stellar mass of the galaxy to be 2$\times$10$^8$ \(\textup{M}_\odot\). WJ004's physical radius, derived from its Kron radius, is 2.65 kpc. Its small size and low stellar mass confirm that WJ004 is a bona fide dwarf galaxy and one of the least massive galaxies known to potentially host an accreting black hole. The archival G band image of WJ004, observed with the Canadian-French-Hawaiian Telescope, is shown in Figure \ref{fig:fig3}.
\subsection{X-ray properties and spectral fitting}
\begin{figure}[t]
\epsscale{1.2}
\plotone{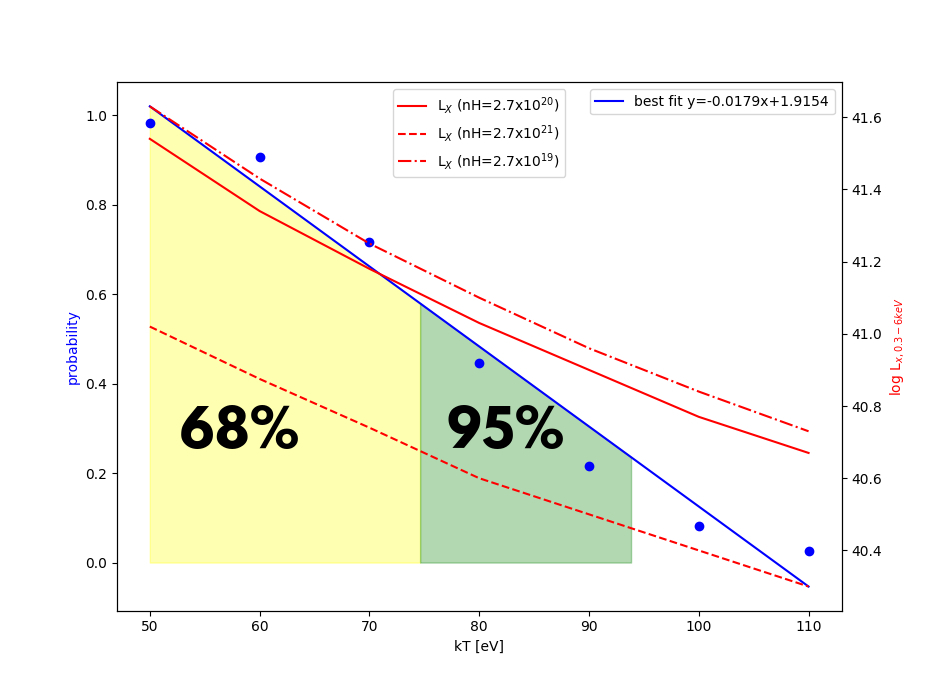}
\caption{Blue dots represent the probability of getting 19 soft photons with respect to the temperature of the blackbody. Solid blue line represents the best fit to the data. Yellow shaded region represents the 68$\%$ confidence interval for the blackbody temperature, while yellow and green regions represent the 95$\%$ confidence interval. Three red lines represent the 0.3-6 keV absorbed X-ray luminosity for 50$<$kT$<$110 eV and for three values of n$_H$.
\label{fig:fig2}}
\end{figure}

The X-ray source is detected only in the soft band, with 19$\pm$6 counts. The statistical significance of the detection is 5.3$\sigma$. No photons above the background are detected in the hard band. To calculate the source's count rate, we calculated the 80\% exposure time-weighted PSF at the location of the X-ray source. The background region was a circular annulus with an inner radius of 7\arcsec and an outer radius of 12\arcsec. The soft and hard band \textit{Chandra} X-ray images are shown in Figure \ref{fig:fig1}.
\newline \indent
In order to translate count rate to flux, we used \textbf{SHERPA}, CIAO's modeling and fitting application. First, we used the CIAO tool \textbf{specextract} to obtain the instrumental response functions, both the auxiliary response file (ARF) and the redistribution matrix file (RMF), for each observation. Then, we used \textbf{combine$\_$spectra} to create summed source counts, response, and background files. Since our source has a low number of counts, no reliable spectral fitting and analysis could be done. In order to overcome this issue, we generated artificial spectra using previously created response files. We specified a source model composed of two components: a photoelectric absorption model and a blackbody emission model, \textbf{xsphabs*xszbbody}. We set the equivalent hydrogen column parameter (n$_H$) to the known Galactic value of 2.7 $\times$ 10$^{20}$ cm$^{-2}$ \citep{2005A&A...440..775K}, and varied the blackbody temperature parameter (kT) from 50 to 110 eV, with increments of 10 eV. Then, we calculated the probability of getting 19 photons under 1 keV and none above 1 keV for each spectrum, and we plotted probabilities versus kT. As expected, lower blackbody temperatures resulted in higher probabilities and higher temperatures resulted in lower probabilities. In the next step we performed the best fit of the data. For simplicity, we used a linear fit of the form Ax+B, where A and B are -0.0179 and 1.9154, respectively. The integration of the linear best fit form, where limits of integration are the lowest and highest temperature, gives us the total area under the best fit curve. The same integration but with two provisional temperatures as limits gives us the area under the curve bounded by those two temperatures. The ratio of that area to total area gives us the probability for spectra being in that range of temperatures. Using this method we found that the spectrum consisting of 19 photons under 1 keV is 68$\%$ likely to originate from a blackbody model with temperature kT$<$74 eV, and 95$\%$ likely to originate from a blackbody model with temperature kT$<$94 eV. We calculated the X-ray luminosity for each spectrum in order to examine how the change of temperature affects the luminosity. We found that at the 68$\%$ confidence level the X-ray luminosity is L$_{X,0.3-6}>$10$^{41}$ erg s$^{-1}$, and at the 95$\%$ confidence level L$_{X,0.3-6}>$7$\times$10$^{40}$ erg s$^{-1}$. In order to further examine how n$_H$ affects luminosity, we performed the same sets of simulations but with n$_H$=2.7 $\times$ 10$^{21}$ cm$^{-2}$ and n$_H$=2.7 $\times$ 10$^{19}$ cm$^{-2}$. We found that the change of n$_H$ across three orders of magnitude does not significantly affect the luminosity. In all cases luminosity stays well above 10$^{40}$ erg s$^{-1}$. The summary of this procedure is shown in Figure \ref{fig:fig2}. \newline \indent
The previous method takes into account only soft photons and assumes no emission above 1 keV. We found that upper limits on hard counts at the 68$\%$, 95.45$\%$ and 99.7$\%$ confidence levels are $<$5.7, $<$13.2 and $<$22.2, respectively. The upper limits were estimated using the CIAO tool aprates. For each upper limit we calculated corresponding upper limits on hardness ratio (HR), defined as HR=$\frac{H-S}{H+S}$, where H and S are numbers of hard and soft photons, respectively. We then created artificial spectra using the \textbf{xsphabs*xszbbody} model with matching hardness ratio. We found that in the 3$\sigma$ upper limit on hard counts, the resulting upper limit on blackbody temperature was kT$<$187 eV, and lower limit on luminosity of L$_{X}>$3.4$\times$10$^{40}$ erg s$^{-1}$. We repeated the same procedure, but using a different model, a photoelectric absorption and power law photon spectrum, xsphabs*xspowerlaw. We found that in the 3$\sigma$ hard counts upper limit the spectrum can be reproduced with the choice of photon index of $\Gamma>$3.9, and resulting X-ray luminosity is L$_X>$7$\times$10$^{40}$ erg s$^{-1}$. At the 2$\sigma$ and 1$\sigma$ level the spectrum becomes considerably softer. Therefore, we find that even the inclusion of the hard component does not significantly alter our findings, and the spectrum remains soft and luminous. The summary of this procedure is given in Table \ref{tab:2}.
\begin{table}
  \centering
  \begin{tabular}{|c|c|c|c|c|c|c|}
  \hline
    (1) & (2) & (3) &(4)&(5&(6)&(7)\\
    \hline
    68$\%$  & $<$5.7&$<$--0.54&$<$118&$>$40.7&$>$3.9&$>$40.8\\
    95.5$\%$&$<$13.2&$<$--0.18&$<$153&$>$40.6&$>$4.7&$>$40.9\\
    99.7$\%$ &$<$22.2&$<$0.08&$<$187&$>$40.5&$>$6&$>$41.2\\
    \hline
  \end{tabular}
  \caption{(1) Confidence levels; (2) Upper limits on hard counts; (3) Upper limits on hardness ratio, defined as HR=$\frac{H-S}{H+S}$; (4) Upper limits on blackbody temperatures kT, from xsphabs*xszbbody model; (5) Lower limits on logarithm of X-ray luminosity for model (4); (6) Lower limits on photon index $\Gamma$, from xsphabs*xspowerlaw model; (7) Lower limits on logarithm of X-ray luminosity for model (6).}
  \label{tab:2}
\end{table}
\section{Discussion} \label{sec:discussion}
\subsection{Probability of false association}
The supersoft source is off-center from the galactic center by 1\farcs5, therefore we need to investigate how likely it is for the supersoft source not to be associated with WJ004. Using deep extragalactic X-ray log $N_X$ vs.\ log $S_X$ results \citep{doi:10.1146/annurev.astro.43.051804.102213} to estimate the number of X-ray sources in the 0.5--2 keV band, per square degree of the sky, we calculated that the probability that a background soft X-ray source at the observed flux level is superposed anywhere within 1\farcs5 radius around any of the 220 dwarf galaxies from our sample to be P=3$\times$10$^{-2}$. However, we note that the survey we used provides the number of background sources in 0.5--2 keV band, while we used the 0.3--1 keV band, which means that the derived probability is overestimated. Also, we did not include the criteria that the source is supersoft, i.e., undetected above 1 keV. In order to quantify the false association probability with the supersoft criteria, we ran the \textbf{wavdetect} tool across the whole Abell 85 field of view, in both soft and hard bands. We found that only our source was detected at the $>$3$\sigma$ level in the soft band but remained undetected in the hard band. The probability for that source to be located anywhere within the 1\farcs5 radius around any of the 220 galaxies from our sample is P=2$\times$10$^{-3}$. Furthermore, the optical image does not reveal any faint, background object that could be associated with the supersoft source, other than WJ004. Therefore, we conclude that the supersoft source is likely to be associated with WJ004. However, even in the case of the supersoft source being a background object, this is still an exciting discovery since supersoft X-ray sources are quite rare.
\subsection{Distance to WJ004}
Due to its faint nature, no spectroscopic redshift measurements exist. WJ004 can be found in SDSS DR7 (cross-identified as SDSS J004144.08-091726.7), and its photometric redshift is estimated to be 0.097. However, the photometric redshift error is 0.138, larger than the redshift itself, probably because the galaxy is flagged for having too few good detections, again a consequence of its faintness. \newline \indent
If we assume that WJ004 and the supersoft source are background objects superposed onto Abell 85, then the galaxy's size, mass and luminosity, and luminosity of the supersoft source increase as well. Nevertheless, this case would also be interesting in its own right. Ultraluminous supersoft active galactic nuclei (AGN) are an extremely rare class of objects, with only three known examples \citep{Terashima_2012,Sun_2013,2013MNRAS.433.1764M,2013ApJ...776L..10L}. If sufficiently distant, the luminosity of the supersoft source would become comparable to the luminosities of supersoft AGN. Supersoft AGN do not show emission above 2 keV, while WJ004 supersoft source does not show emission above 1 keV, which means even if it is more distant than initially thought, it would be an extreme example of an already rare class of objects. Also, supersoft AGN,  even though being significantly more luminous, do not necessarily exclude the existence of an IMBH. This is the best demonstrated in the case of supersoft AGN 2XMM J123103.2+110648 \citep{2012ApJ...759L..16H} for which multiple methods for estimating black hole mass suggest borderline IMBH. Also, the supersoft AGN GN069 is believed to be associated with a low-mass black hole, however not as low-mass as IMBHs \citep{2018ApJ...857L..16S}.  \newline \indent
IMBHs are not the only source of supersoft X-ray emission. Many such sources are known in the Milky Way and other nearby galaxies, and they are confirmed to be associated with recurring post-novae events and steadily accreting white dwarfs. However, they are significantly less luminous than the WJ004 supersoft source, with the maximum possible luminosity of 3$\times$10$^{38}$ erg s$^{-1}$ \citep{2006AdSpR..38.2836K}. Assuming that WJ004 and its supersoft source are foreground objects, significantly closer to us, the luminosity of the supersoft source could decrease and fall in the range of post-novae or accreting white dwarfs. We calculated the required distance at which the luminosity decreases from $\approx$ 10$^{41}$ erg s$^{-1}$ to 3$\times$10$^{38}$ erg s$^{-1}$ to be $\approx$13.5 Mpc. At this distance, the absolute magnitude of the galaxy decreases to M$_V$=--7.5. Such faint galaxies do exist, but they are thought to be the oldest and the least chemically evolved stellar systems, heavily dominated by dark matter \citep{doi:10.1146/annurev-astro-091918-104453}. Thus, such a galaxy is not expected to host a highly luminous supersoft source. Supersoft sources usually have luminosities in the range 10$^{36}$ - few$\times$10$^{37}$ erg s$^{-1}$, meaning that WJ004 could be even closer, which would decrease the supersoft source luminosity furthermore. On the other hand, in that case, the galaxy's absolute magnitude would also decrease, which would amplify its old, chemically unevolved nature. For example, assuming that WJ0004 is a nearby Milky Way object, at a distance of $\approx$ 50 kpc (the distance of Milky Way satellite dwarf galaxies), would imply M$_V$=4.65 and luminosity of 4 $\times$ 10$^{33}$ erg s$^{-1}$. Such an object would be too faint to be a dwarf galaxy or even a globular cluster, since it would be $\approx$ 100 times fainter than some of the faintest low-luminosity globular clusters ever discovered \citep{Koposov_2007}. Another possible  scenario  is  that  WINGS  catalog misidentified a star for a galaxy. In the case of WJ004 being a star, assuming the distance of 5 kpc, its V-band absolute magnitude becomes M$_V$=9.6, which  would make it an M1 star. The X-ray luminosity in this case becomes 4 $\times$ 10$^{31}$ erg s$^{-1}$. WJ004's galactic latitude is -72$^{\circ}$, which is significantly away from the galactic plane, and places the ”star” in the galactic halo. Such a star would have to be an old star with limited magnetic activity, and therefore highly unlikely to produce X-ray emission at this level \citep{2017MNRAS.471.1012B}. Also, we find no GAIA objects within 2$\arcsec$  radius. \newline \indent There is a possibility that the observed supersoft X-ray source is not due to a single emitter, but due to a population of accreting white dwarfs and post-novae events. This would allow the galaxy to be at a distance greater than 13.5 Mpc. We compare this scenario to the case of M31, which has 89 detected supersoft sources, of which a majority have luminosities less than 10$^{38}$ erg s$^{-1}$ \citep{Orio_2010}. This means WJ004 would require hundreds or even thousands of supersoft X-ray sources in order to account for the observed luminosity, greatly exceeding the number of supersoft sources in M31, a galaxy thousands of times more massive than WJ004.\newline \indent
Having all this in mind, we conclude that WJ004 and its supersoft X-ray source are highly unlikely to be foreground objects. They can be background objects; however, such a scenario does not necessarily jeopardize the IMBH explanation.

\subsection{Black hole mass}
Observational evidence suggests that the black hole and galaxy's stellar component regulate each other's evolution. This has led to establishing many scaling relations between the two. One such relation states that there is a connection between the stellar mass of the galaxy and black hole mass; more massive black holes tend to reside in more massive galaxies, while the least massive black holes are expected to be found in the least massive galaxies. The WJ004 black hole mass can be estimated using the relation log(M$_{BH}$/\(\textup{M}_\odot\))=7.45$\pm$0.08+(1.05$\pm$0.11)log(M$_\star$ /10$^{11}$ \(\textup{M}_\odot\)) \citep{Reines_2015}, which yields M$_{BH}\approx$40,000 \(\textup{M}_\odot\). However, this result should be taken with caution since the M$_{BH}$-M$_*$ relation is poorly-calibrated in the low-mass regime.\newline \indent
Another method relies on the fact that supersoft emission, and high/soft states, are associated with high mass accretion rates where an accretor will emit near its Eddington luminosity. The derived luminosity was L$_{X}\approx$10$^{41}$ erg s$^{-1}$, and assuming emission at 10\% L$_{Edd}$, and adopting a bolometric correction $k$=L$_{bol}$/L$_X$=5 \citep{2015MNRAS.448.1893M}, implies an Eddington luminosity of L$_{Edd}$=5$\times$10$^{42}$ erg s$^{-1}$. Since the Eddington luminosity is given as L$_{Edd}$=1.4$\times$ 10$^{38}$ M/\(\textup{M}_\odot\) erg s$^{-1}$, it follows that the mass of accretor is M$_{BH}\approx$35,000  \(\textup{M}_\odot\). \newline \indent
Upon comparing WJ004 to X-ray detected black hole candidates in dwarf galaxies from \cite{Lemons_2015} we noted that only four galaxies are less massive than WJ004. However, none of them harbors an X-ray source as luminous as WJ004's X-ray source. Furthermore, none of the X-ray sources from \cite{Lemons_2015} is supersoft. When it comes to ultraluminous supersoft IMBH candidates, mentioned throughout this paper, most of them are associated with star clusters or tidally-stripped nuclei around or inside large galaxies. On the other hand, WJ004 is an isolated dwarf galaxy with an ultraluminous supersoft source, a unique finding of this kind. 
\subsection{Black hole offset}
\cite{Reines_2020} showed using high-resolution radio observations that the majority of massive black holes in dwarf galaxies are off-nuclear. Some of the galaxies show signs of mergers and interactions, which is a probable reason for the offset of their black holes. This is also in accordance with simulations that predict a significant population of wandering massive black holes in dwarf galaxies due to mergers and interactions \citep{2019MNRAS.482.2913B}. Therefore, the observed offset in WJ004 is not unexpected ($\approx$1.8 kpc). \newline \indent 
An important question we need to answer to understand IMBHs better is why the galaxy WJ004 stands out among many other similar galaxies that do not host active, highly accreting IMBHs. We suggest that the galaxy interactions, the reason behind black hole offset, are also responsible for enhanced black hole activity. The idea that galaxy interactions can trigger black hole activity is not new and has been widely explored from theoretical, observational, and computational standpoints. \cite{pr10.1093/mnras/stab2740} showed using hydrodynamical simulations that during galaxy interactions, gravitational and hydrodynamical torques can have a significant influence on angular momentum and mass redistributions, which can lead to inward mass transport onto the black hole influence radius. They also showed that close passages could trigger accretion that approaches the Eddington rate, and as we already mentioned, supersoft emission is expected to be associated with high accretion rates. Observations of interacting galaxies up to $z\leq$1, showed that the link between black hole fueling and close galaxy passages does exist and that interacting galaxies are by at least a factor of three more likely to have an accreting black hole than non-interacting galaxies, and this effect is independent of stellar mass \citep{10.1093/pasj/psx135}. \newline \indent
Another possible explanation behind enhanced black hole activity could be tidal disruption events (TDEs). IMBH-related TDEs should produce thermal-state signatures with high luminosities and supersoft spectra. Recent works suggest that TDEs are one of the most efficient methods to detect IMBHs \citep{2018NatAs...2..656L, Wen_2021}. However, even though our detection exhibits properties that are in line with the TDE scenario, we cannot confirm or refute this hypothesis due to insufficient data.

\section{Conclusions}\label{sec:conc}
The presence of black hole seeds at $z>7$ was introduced in order to explain the existence of SMBHs at very large redshifts. It is challenging to observe the first black hole seeds directly, but those seeds that did not grow to supermassive sizes should be observed in low-redshift low-mass galaxies. We used archival \textit{Chandra} observations to study X-ray sources in dwarf galaxies and search for unique signatures of IMBHs. 
\begin{itemize}
    \item We detected a supersoft X-ray source associated with dwarf galaxy WJ004. The source shows no emission above 1 keV.
    \item The supersoft nature and luminosity of the detected source are in line with an IMBH explanation.
    \item We discussed that it is highly unlikely for the source to be a background/foreground object, a recurring post-nova event or a steadily accreting white dwarf.
        \item We suggest that the reason behind the IMBH's offset, enhanced activity, and supersoft spectrum could be galaxy interactions.
\end{itemize}
\vspace{0.1in}\hspace{0.85in}\normalsize{ACKNOWLEDGMENTS}\vspace{0.1in} \newline 
Authors gratefully acknowledge the use of \textsc{CIAO} software \citep{2006SPIE.6270E..1VF}, as well as the Ned Wright Cosmology Calculator \citep{Wright_2006}. DL is supported by NASA ADAP Grant NNX10AE15G.
\bibliography{sample62}{}
\bibliographystyle{aasjournal}


\end{document}